\documentclass[aps,prl,a4,superscriptaddress,preprintnumbers,showpacs,twoside,twocolumn,floatfix]{revtex4}

\usepackage{bm}
\usepackage{graphicx}
\usepackage{amsfonts}
\usepackage{amsmath}
\usepackage{amssymb}
\usepackage{amsthm}
\usepackage{bbm}
\usepackage{setspace}
\usepackage{times}
\usepackage{color} 

\theoremstyle{plain}

\begin{document}

\title{Pairing, crystallization and string correlations of mass-imbalanced atomic mixtures in one-dimensional optical lattices}

\author{Tommaso Roscilde}
\affiliation{Laboratoire de Physique, CNRS UMR 5672, Ecole Normale Sup\'erieure de Lyon, Universit\'e de Lyon, 46 All\'ee d'Italie, 
Lyon, F-69364, France}
\author{Cristian Degli Esposti Boschi}
\affiliation{CNR-IMM, Sezione di Bologna, via Gobetti 101, I-40129, Bologna, Italy}
\author{Marcello Dalmonte}
\affiliation{Dipartimento di Fisica and INFN, Universit\`a di Bologna, via Irnerio 46, I-40126, Bologna, Italy}
\date{\today}

\begin{abstract}
We numerically determine the very rich phase diagram of mass-imbalanced binary mixtures 
of hardcore bosons (or equivalently -- fermions, or hardcore-Bose/Fermi mixtures) loaded 
in one-dimensional optical lattices. Focusing on commensurate fillings away from half filling, 
we find a strong asymmetry between attractive and repulsive interactions. Attraction 
is found to always lead to pairing, associated with a spin gap, and to pair crystallization for
very strong mass imbalance. In the repulsive case the two atomic components remain instead 
fully gapless over a large parameter range; only a very strong 
mass imbalance leads to the opening of a spin gap. The spin-gap phase is the precursor of a crystalline phase occurring
for an even stronger mass imbalance. The fundamental asymmetry of the phase diagram is at odds with recent 
theoretical predictions, and can be tested directly via time-of-flight experiments on trapped
cold atoms.  
\end{abstract}






\maketitle

  One-dimensional quantum liquids occupy a special place in the context of 
 quantum many-body systems: indeed interactions of any strength lead to 
 quantum fluctuations as strong as to discard Bose condensation for bosons
 and the Fermi liquid picture for fermions down to zero temperature. For sufficiently
 weak interactions a new unifying paradigm of the so-called Tomonaga-Luttinger 
 liquids (TLL) emerges \cite{Giamarchi04}, characterized by the fact that all 
 elementary excitations are gapless, and both diagonal and 
 off-diagonal correlations decay algebraically with the distance. Recent advances
 in the trapping of ultracold atoms in optical lattices allow
 to realize one-dimensional quantum liquids in a highly flexible way, with the possibility
 of fully controlling the statistics and the interaction strength \cite{Blochetal08}. 
A series of recent experiments has demonstrated the physics of one-dimensional Bose gases  
with strong interactions up to the hardcore (or Tonks-Girardeau) limit 
\cite{Kinoshitaetal04}. A special role in the context of one-dimensional systems is played by 
binary mixtures, either bosonic, fermionic, or Bose-Fermi ones, for which TLL theory predicts 
the separation of spin and charge modes \cite{Giamarchi04}. In the case of particles with 
equal masses and repulsive short-range interactions, both charge and
spin sectors can be gapless, and one recovers an effective picture of two decoupled TLLs. 
Such a picture can be made unstable via several mechanisms: via Mott localization
in presence of an underlying lattice and for integer total filling; via localization into
a true (long-range-ordered) crystal (TC) in presence of a strong off-site repulsion; via phase separation; or 
via the formation of bound states (\emph{e.g.} Cooper pairs for attractive interactions) 
leading to the appearance of  a spin gap.  
\begin{figure}[tb]
\includegraphics[width=85mm]{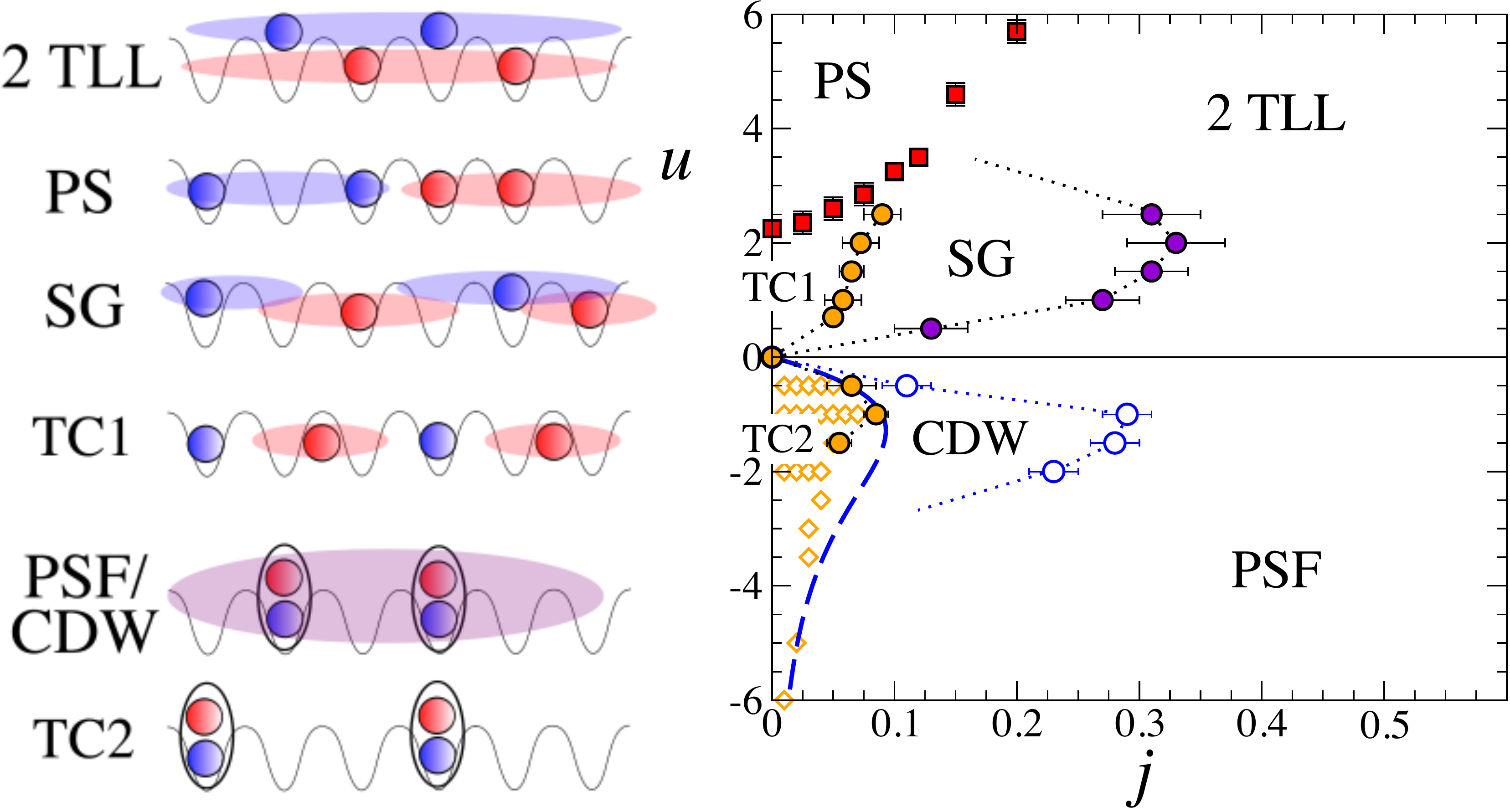}
\caption{Phase diagram of a mass-imbalanced atomic mixture with filling $n_a = n_b = 1/3$.  
The phase boundaries are determined via QMC; the open diamonds indicate points in parameter space for which 
DMRG finds a fully gapped crystal phase TC2. The dashed line indicate the 
FK gap $\Delta_{\rm k}/J_a$ for kink-antikink pairs. The left panel presents a sketch of the phases:
2 TLL = 2 Tomonaga-Luttinger liquids; PS = phase separated; SG = spin gap; PSF = pair superfluid; 
CDW = algebraic charge density wave; TC1 and TC2: true crystals.}
\label{f.PHD}
\vspace*{-.5cm}
\end{figure}

Here we show that the TLL picture undergoes a complex series of instabilities 
in binary mixtures with \emph{mass imbalance} between the two species. 
We focus here on quantum particles on a lattice with intraspecies hardcore repulsion and on-site
interspecies interactions, describing 
at the same time spin-1/2 fermions, spin-1/2 hardcore bosons, and mixtures of hardcore
bosons and spinless fermions.  
  The system Hamiltonian reads 
 \begin{equation}
 \null\hspace*{-.2cm}
 {\cal H} =  \sum_{i} \left[-(J_a a^{\dagger}_i a_{i+1} + J_b b^{\dagger}_i b_{i+1} + \text{h.c.})  
 + U n_{i,a} n_{i,b}\right]
 \end{equation}
 in which $a$ and $b$ correspond to the two atomic species. Mass imbalance
 is controlled by the ratio $j = J_b/J_a$ and interaction by the ratio $u=U/J_a$. 
 As far as the spectrum and the diagonal observables are concerned, 
 we do not need to specify the statistics of the $a$ and $b$ operators; for what
 concerns off-diagonal observables, unless otherwise specified we will 
 refer explicitly to hardcore bosons, satisfying bosonic commutation relations off-site
 and fermionic anticommutation relations on-site. The denomination of the 
 many-body phases will also be mostly inspired by the case of hardcore-boson mixtures. 
  
  We consider both repulsive ($u>0$) and attractive ($u<0$) interactions, and we 
 focus on the case of equal densities away from half filling, $n_a = n_b = 1/p \neq
  1/2$, $p\in\mathbb{N}$. 
 The system with equal masses, $j=1$, is integrable \cite{Hassleretal05},
 and features spin-charge separation into two TLL in the repulsive case; in the 
 attractive one, an $a$-$b$ bound state appears, associated with the opening 
 of a spin gap, and only the charge sector remains gapless, giving rise to a paired 
 superfluid (PSF) phase.      
 The case of mass imbalance has been studied recently by bosonization 
 \cite{Cazalillaetal05, Mathey07, Wangetal07}, and numerically via the density-matrix
 renormalization group (DMRG) and related approaches \cite{RizziI07, Wangetal09, Barbieroetal10}. 

 Making use of numerically exact methods, here we determine comprehensively
 the rich phase diagram of the system with mass imbalance, $j<1$, as shown in Fig.~\ref{f.PHD}. 
 The latter figure refers to the case $n=1/3$, but the qualitative features are 
 generic for commensurate fillings. Our main findings are the following: 
 1) in the attractive case, the PSF phase is found to persist up to strong  
 imbalance, at which the system becomes unstable to the formation of a TC of pairs. 
 2) in the repulsive case, on the other hand, the double TLL 
 of the mass-balanced case survives up to a large imbalance, at which
 two different instabilities appear. For sufficiently large repulsion, the two species
 phase-separate. For weaker repulsion, first a spin-gap opens, associated  
 with fluctuating magnetic order captured by string correlations; 
 this phase is the precursor of a TC phase with pinned $b$ particles, 
 and $a$ particles localized in the interstitial regions. 
 
 Before discussing the derivation of the phase diagram, we point out
 that the physics described in this paper is largely accessible to current experimental
 setups on ultracold mixtures in one-dimensional optical lattices.  The mass imbalance
 can be realized with heteronuclear mixtures (\emph{e.g.} $^{40}$K-$^6$Li, for the fermionic
 case, $^{41}$K-$^{87}$Rb for the bosonic case, and $^{40}$K-$^{87}$Rb for the 
 Bose-Fermi case \cite{Chinetal10}),  and with homonuclear mixtures in different 
 hyperfine states, and it can be continuously tuned by using lasers with a
 wavelength close to the magic value or to an atomic resonance for one of the two species. 
 The interspecies interaction can be tuned by Feshbach resonances, as widely
 demonstrated in the recent literature \cite{Chinetal10}. 
     
   The phase diagram in Fig.~\ref{f.PHD} is the result of a joint numerical study based on 
   quantum Monte Carlo (QMC) and DMRG\cite{White92}. Our QMC calculations are based on a canonical
   formulation \cite{Roscilde08} of the Stochastic Series Expansion approach with directed loops 
   \cite{Syljuasen03}, applied to chains with up to $L=150$ sites with periodic boundary conditions, 
   and at temperatures $\beta J_b = L$ capturing the $T=0$ physics for both species. Our DMRG 
   calculations apply to chains with up to $L=144$ with open boundary conditions and retaining
   up to $M=1400$ states. 

   We start our discussion by the attractive case. For a very broad range of mass imbalance, the system 
   displays a PSF phase, characterized by quasi-condensation of bound $a$-$b$ pairs, giving rise
   to an algebraic decay of the pairing correlation function \cite{Giamarchi04}, 
   $G_{ab}(r) = \langle a^{\dagger}_i b^{\dagger}_i b_{i+r} a_{i+r} \rangle \sim r^{-1/K_\rho}$;
  density-density correlations are also decaying algebraically as 
  $C_{\rho}(r) = \langle n_i n_{i+r} \rangle - \langle n_i \rangle \langle n_{i+r}\rangle  \approx -\frac{K_{\rho}}{\pi^2 r^2} +  A \frac{\cos(2\pi n r)}{ r^{K_\rho+K_{\sigma}}}$;
  here, $n_i = n_{i,a}+ n_{i,b}$, and $K_{\rho}$ is the charge {TLL parameter}. For equal masses, 
  $K_\rho > 1$ \cite{Giamarchi04} for all $u < 0$, so that the dominant correlations
  are the pairing ones. Moreover $K_{\sigma}=0$ due to the presence of the spin gap (see below).  
  We extract the Luttinger exponent from the slope of the density
  structure factor at $q\to 0$, $S_{\rho}(q) =  \sum_r \exp(iqr) C_{\rho}(r) \approx K_{\rho}~q/\pi$, and 
  we find that mass imbalance leads to a reduction of $K_\rho$, consistently with what
  observed for other fillings in Ref.~\cite{Barbieroetal10}.  For large mass imbalance,  
  $K_{\rho}$ becomes smaller than one: this corresponds to the loss of quasi-condensation, 
  in favor of a quasi-solid phase (or charge density wave, CDW), with dominant density
  correlations. This phase is the precursor of a quantum phase transition to a TC of pairs --
  phase TC2 of Fig.~\ref{f.PHD} -- 
  with the onset of long-range density order at wavevector $Q = 2 \pi n$. We determine the extent 
  of the TC phase via QMC by determining the mass imbalance at which $S_{\rho}(Q)$ starts 
  diverging linearly with system size; and by DMRG detecting the onset of the exponential 
  decrease of $G_{ab}$ and $C_{\rho}$, marking the opening of a charge gap \cite{SuppMat}.     
   The TC instability is well understood coming from the Falicov-Kimball (FK) limit, $J_b =0$. 
  In this limit, which reduces to a 1D lattice gas in a static potential, we find that the ground state corresponds
  to the TC of pairs for all values $u<0$. The gap $\Delta_{\rm k}$ to the formation of kink-antikink pairs in the TC
  is found to be a non-monotic function of $|u|$, displaying an intermediate maximum. We observe that, for small $|u|$,  
  the boundary of the crystalline region follows closely the locus at which the gap $\Delta_{\rm k}$ equals $J_b$, and 
  it has a re-entrant shape mimicking the non-monotonic behavior of the gap as a function of $u$. 
  This suggests that the quantum melting 
  transition corresponds to a condensation of kink-antikink pairs in the ground state. 
 \begin{figure}[tb]
\includegraphics[width=76mm]{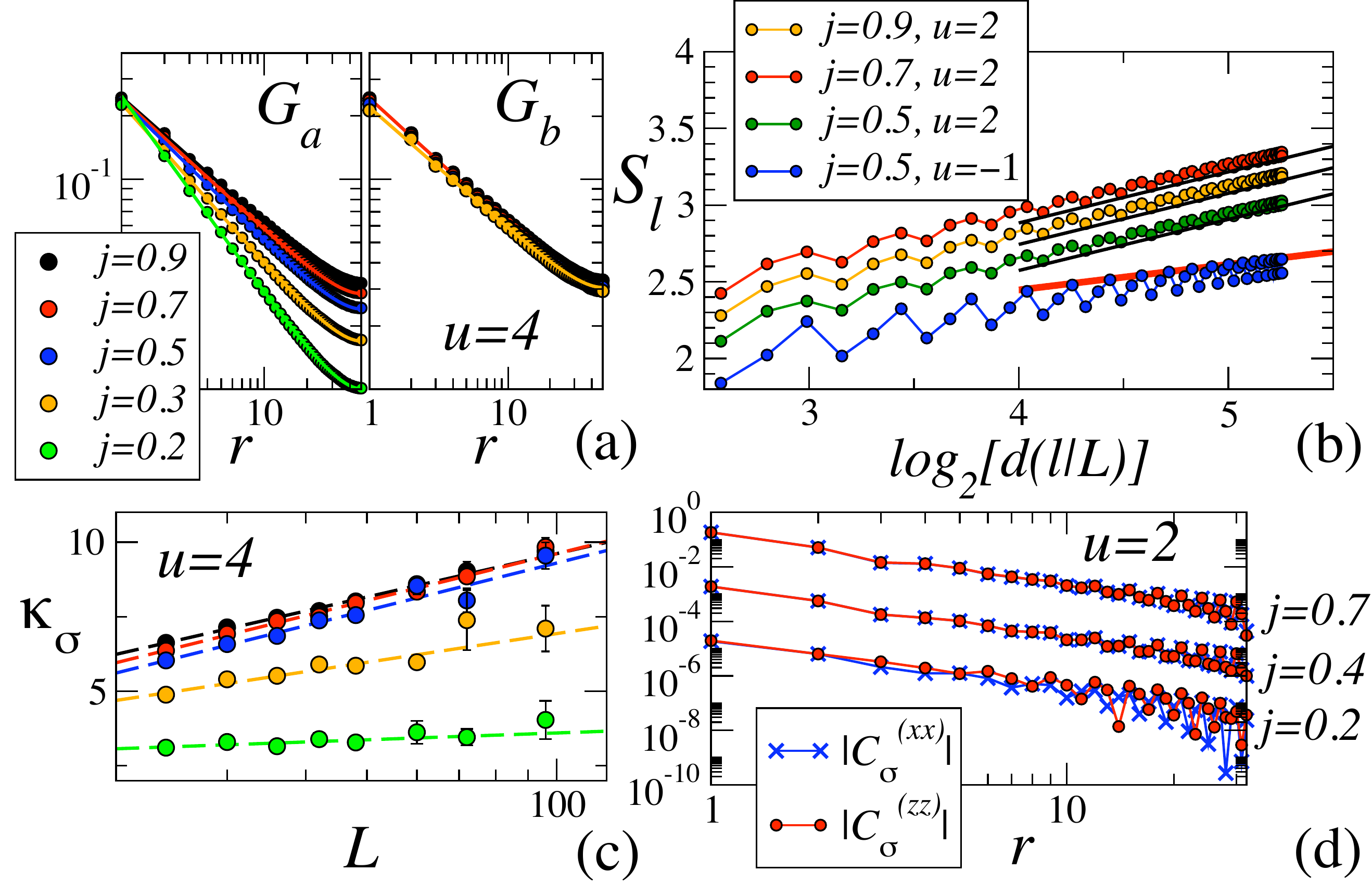}
\caption{
Evidences for the absence of spin gap for moderate mass imbalance. 
(a) Green's function for both species (solid lines are fits
to $A_{a(b)} d(r|L)^{-1/(2K_{a(b)})}$, $L=96$); (b) $\kappa_\sigma$
coefficient (dashed lines are linear fits); all data are for $u=4$;
(c) Block entropy $S_l$ for 
chains with $L=120$ sites (the curves for $j=0.5$ and $j=0.7$ are
shifted to improve readability); thick lines are a reference for $c=2$
 (black) and $c=1$ conformal field theories (red); (d) Fermionic spin-spin correlation functions, calculated
 via DMRG between the sites $L/3$ and $2L/3$ of a chain with $L=96$. The data for $j=0.4$ and $j=0.2$
 have been rescaled to improve readability (by a factor $10^{-2}$ and $10^{-4}$, respectively).}  
 \vspace*{-.4cm}
\label{f.2TLL}
\end{figure}

   The repulsive side of the phase diagram is more complex. Combining bosonization with 
   renormalization group calculations up to two loops,  Refs.~\onlinecite{Cazalillaetal05, Mathey07} 
   conclude that a spin gap should open for
    any infinitesimal mass imbalance  as $\Delta_s \approx \Lambda \exp(-A'/|J_a-J_b|)$ for $J_b \lesssim J_a$, 
    where $\Lambda \sim J_a, J_b$. On the contrary, Ref.~\onlinecite{Wangetal07}, also based upon bosonization, 
    concludes that the spin gap is absent in the repulsive case. 
   All our numerical findings point toward
   the persistence of a fully gapless TLL behavior for both the charge \emph{and} spin sector over a dominant 
   portion of the repulsive phase diagram. Our conclusion is based on a number of crossed evidences.
   First of all, we observe that the one body correlation functions $G_a(r) = \langle a^\dagger_i a_{i+r} \rangle$
   and $G_b(r) = \langle b^\dagger_i b_{i+r} \rangle$  can be very well 
   fitted with the simple power-law form $G_{a(b)}(r) = A_{a(b)} d(r|L)^{-1/(2K_{a(b)})}$, where
   $d(r|L) = L |\sin(\pi r/L)|/\pi$ is the conformal distance (see Fig.~\ref{f.2TLL}(a)).  In particular we find that,
  for weak and moderate repulsions, $K_{a(b)} > 0.5$, which implies that the momentum
  distribution $n_{a(b)}(q) = \sum_r \exp(iqr) G_{a(b)}(r)$ displays a quasi-condensation 
  divergent peak at $q=0$, to be detected in time-of-flight experiments (see below). 
 Moreover we exploit the fact that an explicit counting of the number of gapless degrees of 
   freedom in the system comes from the central charge $c$ of the conformal field theory corresponding
   to our model of interest. This quantity can be directly extracted via DMRG, using the fundamental
   result that the entanglement entropy (EE) of a boundary block of the system grows 
   with the size $l$ of the block \cite{CalabreseC04} as
   $S_l = -{\rm Tr}(\rho_l \log_2 \rho_l) \approx  \frac{c}{6} \log_2[d(l|L)] + \mathcal{O}(1/l)+ {\rm const}$,
   where $\rho_l$ is the reduced density matrix of the boundary block, and 
   open boundary conditions are employed. Fig.~\ref{f.2TLL}(b) (see also Ref.~\cite{SuppMat})
   shows that the scaling of the EE is fully consistent with $c=2$ in the repulsive regime, providing further 
   evidence for the fact that the TLL has two gapless components even for a significant
   mass imbalance.  
   
Finally, using QMC we gain further insight into the gapless phase by 
investigating the spin-spin correlation function  
$C^{(zz)}_{\sigma}(r) = \langle S^z_i S^z_{i+r} \rangle  \approx -\frac{K_{\sigma}}{\pi^2 r^2} +  B \frac{\cos(2\pi n r)}{ r^{K_\rho+K_\sigma}}$;
  where $S^z_i = (n_{i,a}- n_{i,b})$, and $K_{\sigma}$ is the spin TLL parameter; in absence
 of a spin gap, $K_\sigma \geq 1$.  We extract the Luttinger exponent from the low-$q$ behavior of 
 the spin structure factor, $S_\sigma(q) \approx K_\sigma ~q/\pi$, giving the finite-size
 estimate $K_\sigma(L) = [(L/2) S_\sigma(2\pi/L;L) + (L/4) S_\sigma(4\pi/L;L)]/2$. 
 At the SU(2) invariant point $j=1$, $K_\sigma(L)$ is known to obey the 
 Kosterlitz-Thouless (KT) critical scaling  $K_\sigma(L) \approx 1 + [a \log(L/L_0)]^{-1}$ \cite{Giamarchi04}, 
 which implies that the quantity $\kappa_\sigma(L) = [K_\sigma(L) - 1]^{-1}$ scales
 linearly with $\log(L)$. Quite remarkably, this scaling law remains valid over a broad range of
 $j<1$ values, and even for strong inter-species coupling, as shown in Fig.~\ref{f.2TLL}(c).
 This means that, even if the SU(2) symmetry is broken in the Hamiltonian, 
 it appears to be restored in the ground-state of the system, namely that a
 moderate mass imbalance is an \emph{irrelevant} perturbation and that, under
 a renormalization group treatment, the system flows to the SU(2)-symmetric 
 fixed point $K_\sigma(L\to\infty)\to 1$ over a broad range of $j$ values. 
 This picture is further corroborated by comparing the 
 longitudinal spin-spin correlations $C^{(zz)}_{\sigma}(r)$ 
 and the transverse ones in the \emph{fermionic case}, $C^{(xx)}_{\sigma}(r) = \langle S^x_i S^x_{i+r} \rangle$, 
 where $S_i^x = a_i^{\dagger} b_i +  b_i^{\dagger} a_i$; the two correlation functions
 show the same algebraic decay over a broad range of $j$ values 
, and they even appear to coincide at short range, as shown in Fig.~\ref{f.2TLL}(d). 
 \begin{figure}[tb]
\includegraphics[width=76mm]{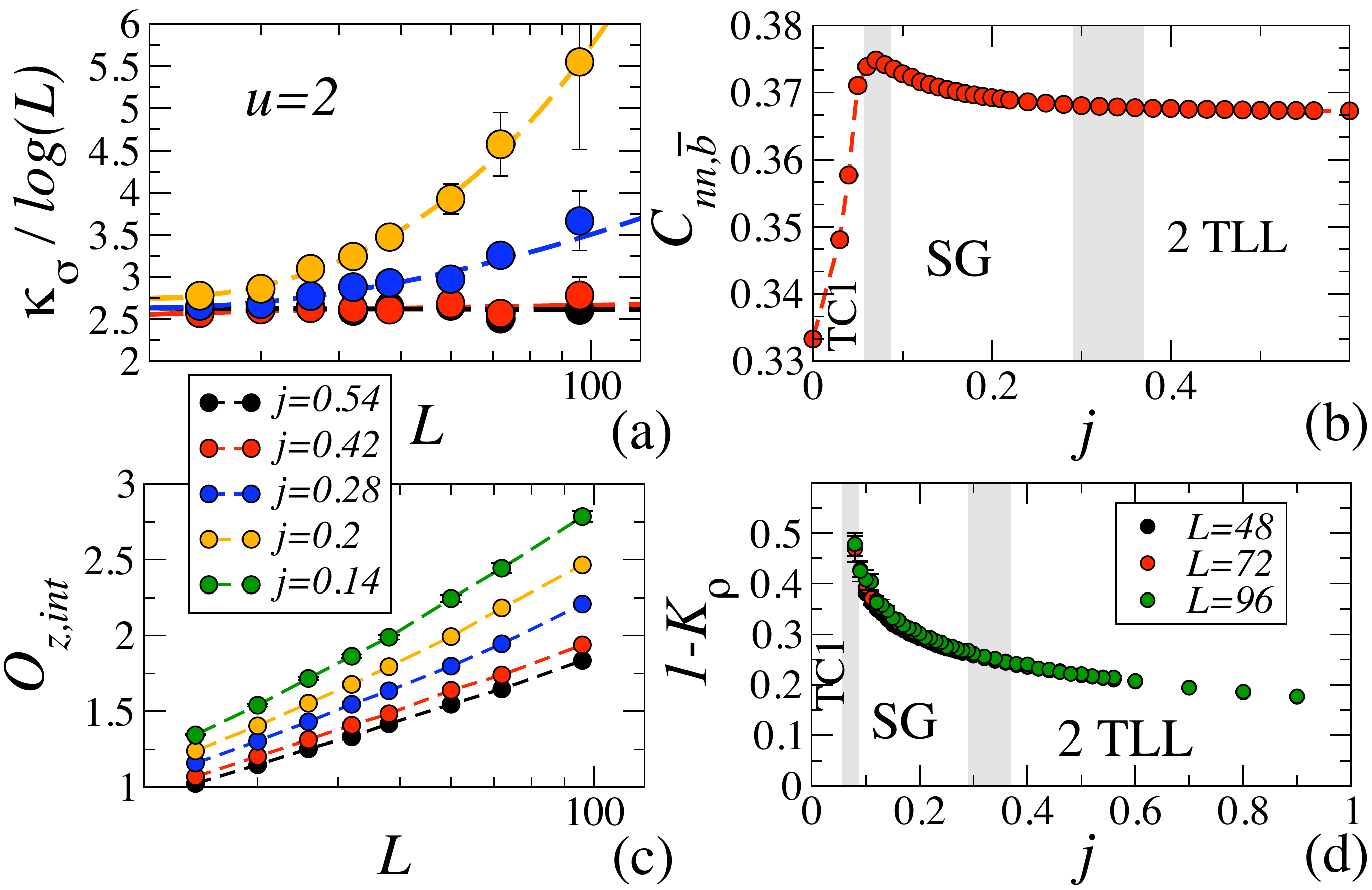}
\caption{(a) Breakdown of SU(2)-invariant scaling of $\kappa_\sigma$ for 
$u=2$ and decreasing $j$;
dashed lines are fits to $a_1/\log(L)+a_2\log(L)^{a_3}$;
(b) $b$-hole density correlator for nearest neighbors, $L=30$; the shaded areas mark the
transition regions among the various phases; (c) String structure factor
as a function of system size (same parameters as in (a)); (d) Decay exponent 
of the string correlation function as function of $j$.}
\null\vspace*{-.8cm}
\label{f.SG}
\end{figure}

   On the other hand, a careful study of the behavior of the system shows that 
   the two-component TLL becomes indeed unstable to the formation of a spin gap (SG) for
   \emph{strong} mass imbalance ($j \lesssim  1/3$ for $n = 1/3$). 
   The opening of a spin gap leaves a clear signature in the Luttinger exponent 
   $K_\sigma$, which stops flowing to the SU(2) symmetric value: this is seen in the fact
   that $\kappa_\sigma$ begins to scale \emph{faster} than linearly in $\log(L)$,
   which is a manifestation of the fact that $K_\sigma(L)$ flows to values 
   smaller than 1, as shown in Fig.~\ref{f.SG}(a). We use this criterion to locate
   the opening of the spin gap in Fig.~\ref{f.PHD}. Moreover the breaking of SU(2)
   symmetric scaling is also seen in the spin-spin correlation functions, which 
   acquire a different behavior at short and long distances (see Fig.~\ref{f.2TLL}(d)).
  The opening of a spin gap signals the formation of a bound state of some sort:
  but what could this bound state be in a purely repulsive system? We argue that
  a bound state can appear between the $a$ particles and the \emph{holes}
  of the $b$ species, which are in a commensurate proportion $m=(1-n)/n$ to the $a$
  particles. As shown in Ref.~\onlinecite{Burovskietal09}, a sufficiently strong 
  mass imbalance can indeed bind together particle composites in 1D quantum 
  fluids - in the specific case of $n=1/3$ we have trimers formed by one $a$
  particle and two $b$ holes ($\bar{b}$). For strong enough repulsion ($u\gtrsim 1$) 
  the apparition of $a$-$2\bar{b}$ composites
  is seen in the nearest-neighbor density correlations of the $\bar{b}$ holes, 
  $C_{nn,\bar{b}} = \langle (1- n_{i,b})(1-n_{i+1,b}) \rangle$, which is enhanced
  for increasing mass-imbalance, as seen in Fig.~\ref{f.SG}(b). 
  \begin{figure}[tb!]
\mbox{
\includegraphics[width=40mm]{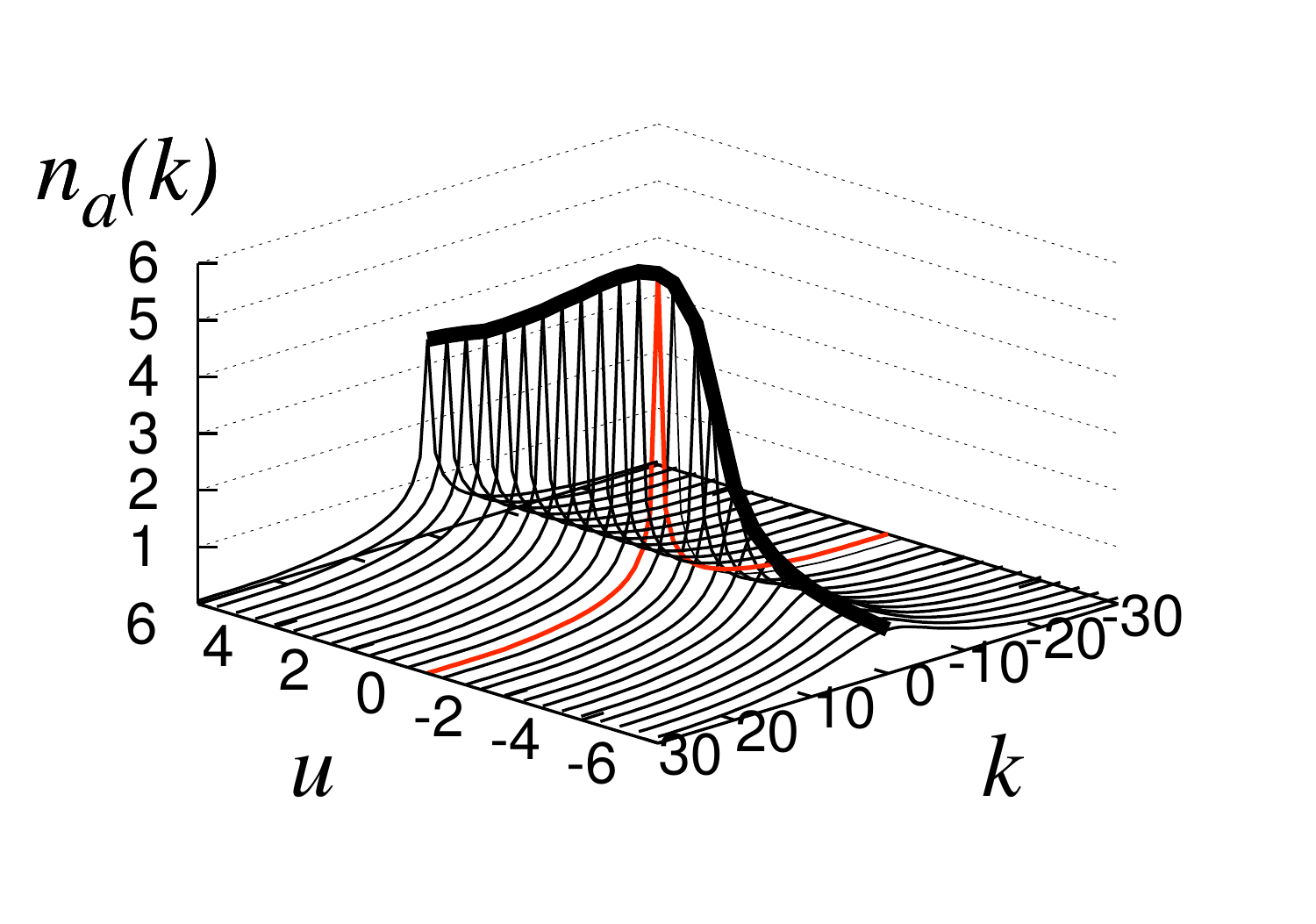}
\includegraphics[width=40mm]{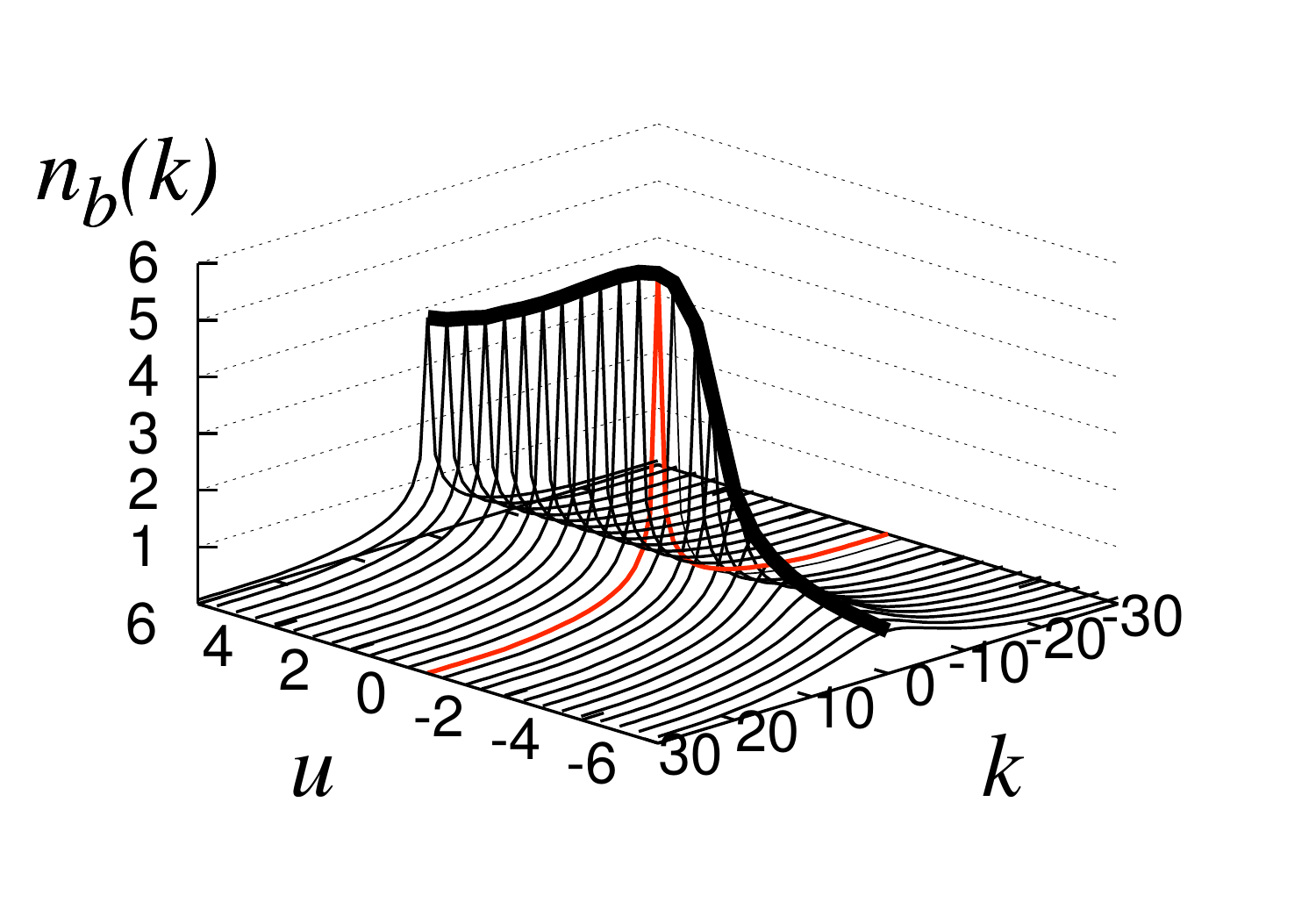}}
\mbox{
\includegraphics[width=40mm]{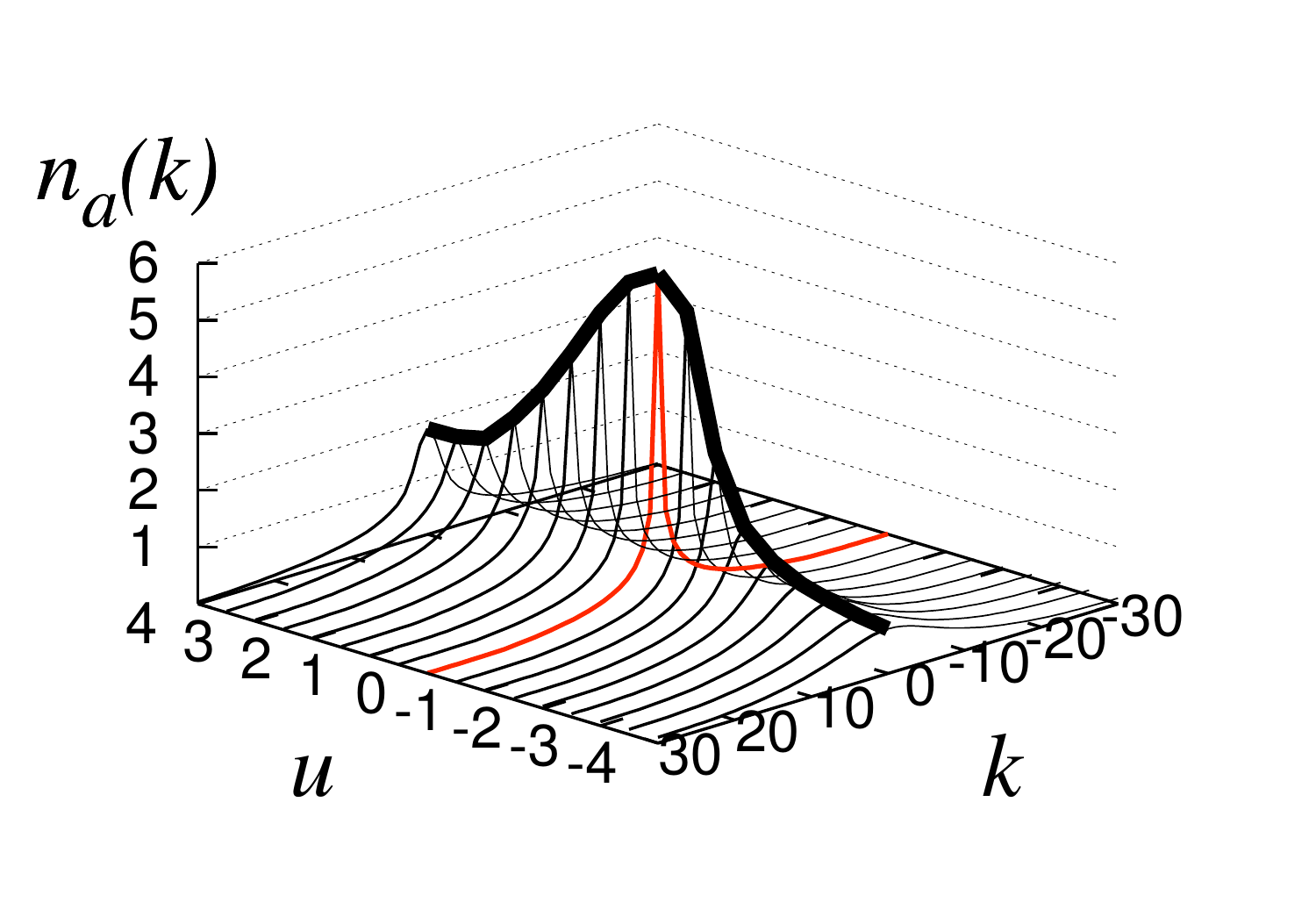}
\includegraphics[width=40mm]{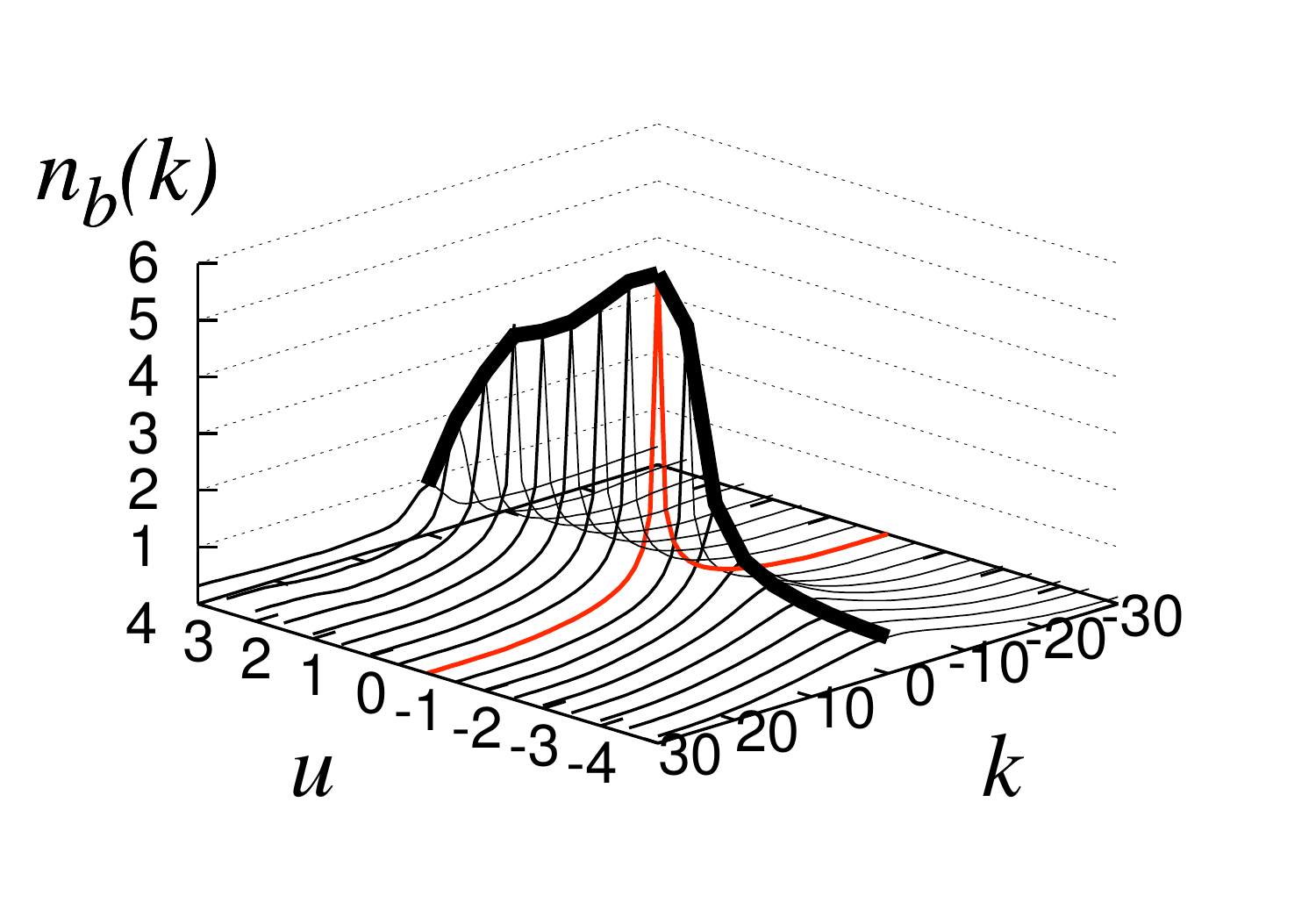}}
\mbox{
\includegraphics[width=40mm]{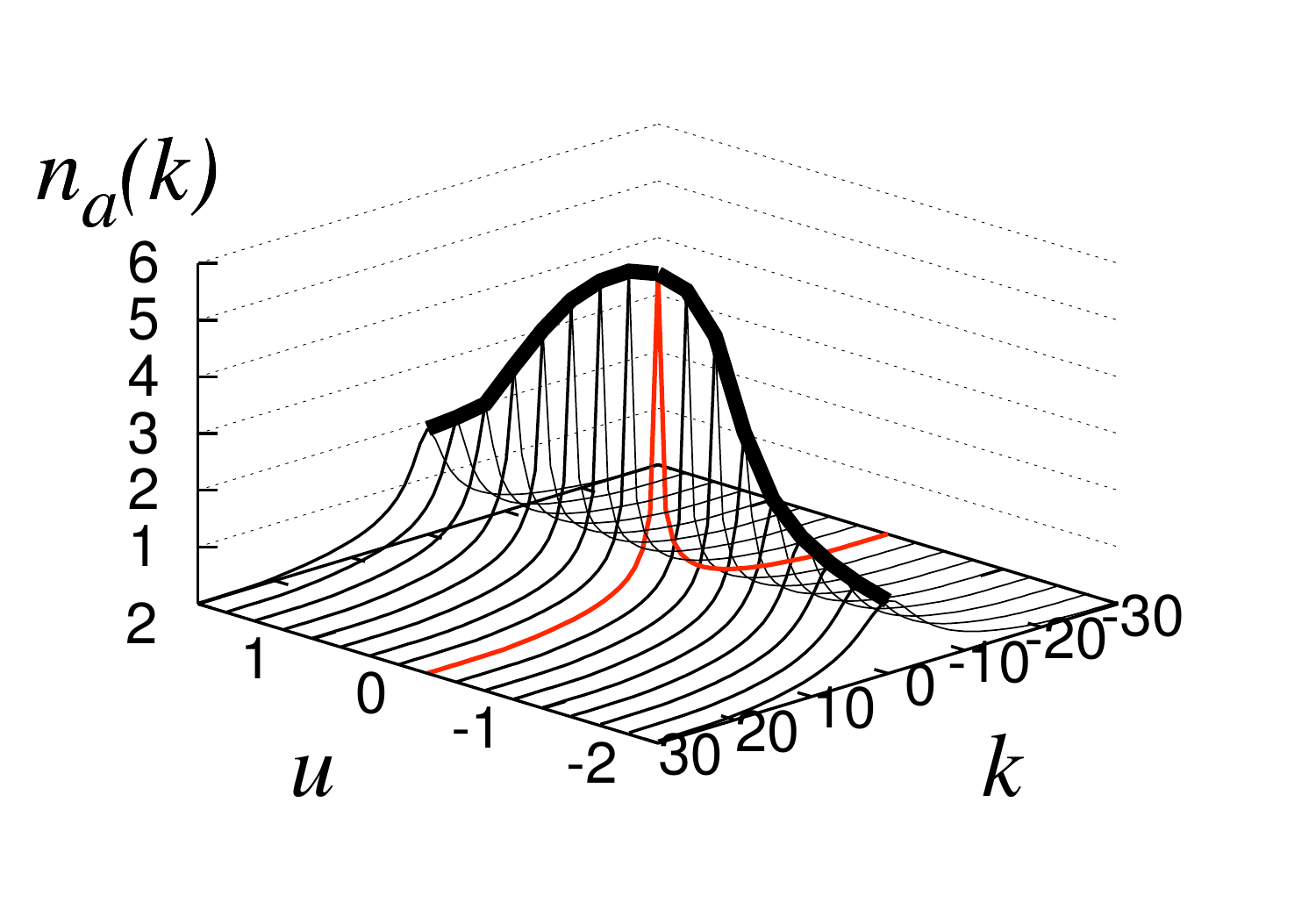}
\includegraphics[width=40mm]{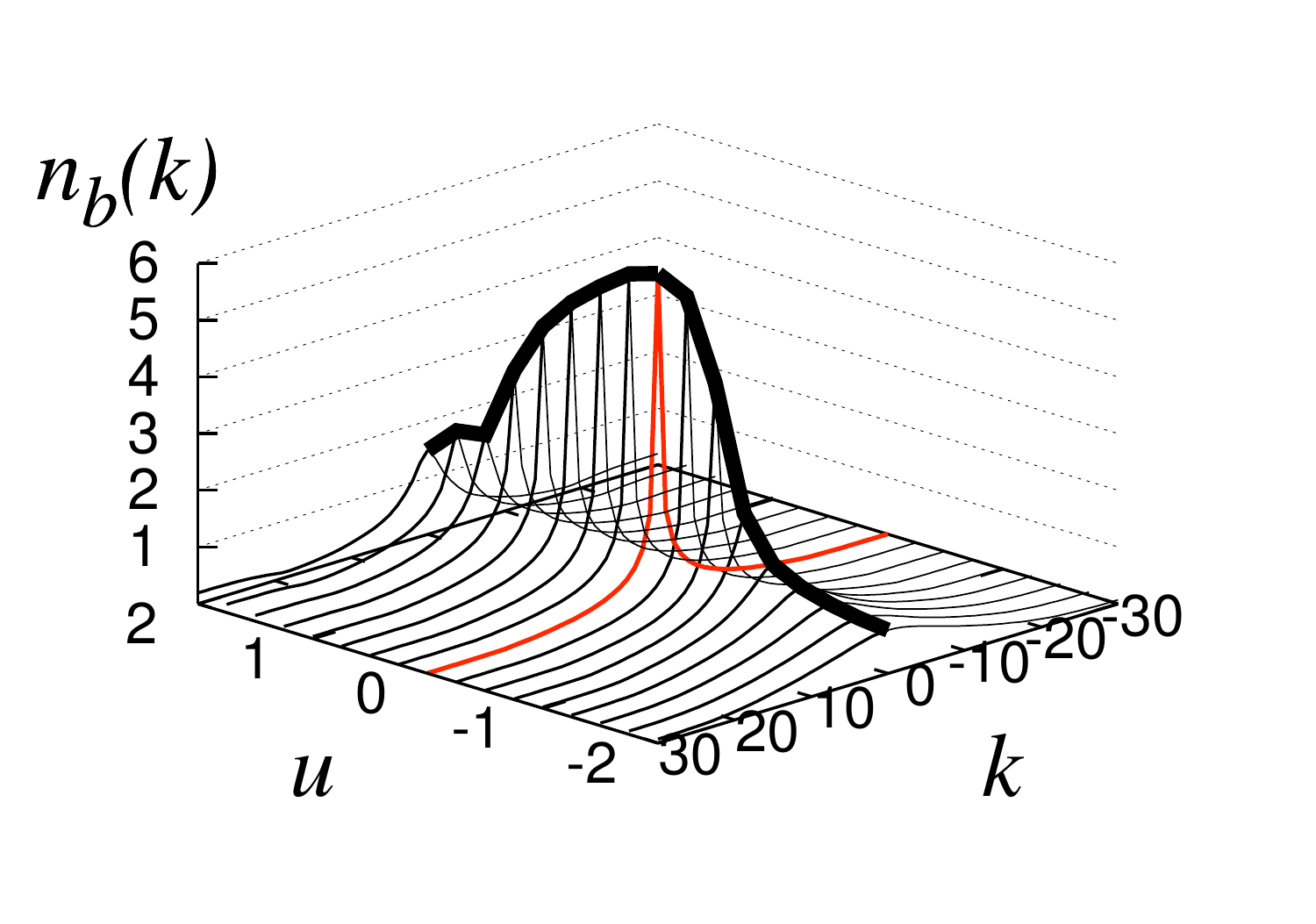}}
\caption{Evolution of the momentum distribution of both species
on a $L=60$ chain,  
for interactions spanning the repulsive and the attractive side
at fixed mass imbalance. From top to bottom: $j = 2/3, 1/10, 1/20$.}
\label{f.nk}
\vspace{-.4cm}
\end{figure}

  The $a$-$2\bar{b}$ composites have mutual 
  repulsive interactions, as observed in Ref.~\onlinecite{Keilmannetal09}, 
  so that their binding favors configurations of the type $..a0ba0b0a0b0ab...$  in which 
  the $a$ and $b$ particles alternate spatially. Such configurations, albeit lacking
  long-range density and magnetic order, contain strong non-local correlations that can be 
  captured using a \emph{string correlation function}\cite{denNijsR89} 
   $O_z(r) = - \left \langle S^z_i~ e^{i\pi\sum_{j=i+1}^{i+r-1} S^z_j} ~S^z_{i+r} \right\rangle$~.   
   As shown in Fig.~\ref{f.SG}(c-d), string correlations are significantly enhanced when 
   entering the SG phase:  indeed the string structure factor, $O_{z,int} = \sum_r O_z(r)$,
   shows a more pronounced divergence with system size, which signals the slower 
   decay of $O_z(r)$. As shown in Ref.~\onlinecite{SuppMat}, in presence of spin-charge 
   separation the dominant decay of the string correlations is governed solely by the 
   $K_{\rho}$ exponent, $O_z(r)\sim r^{-K_\rho}$; therefore $K_\rho$ controls the 
   divergence of  $O_{z,int} \sim L^{1-K_{\rho}}$, and in Fig.~\ref{f.SG}(d) we observe that 
   the $1-K_{\rho}$ exponent (extracted from the low-$q$ behavior of $S_{\rho}(q)$) is
   significantly enhanced in the SG phase. Given that both the spin and charge structure factor  
   exhibit a peak whose divergence with system size is hindered by the slow
   size-scaling of the divergence exponent $1- K_s - K_\rho$ \cite{SuppMat}, we find that the 
   string correlations best characterize the ``fluid" magnetic order of the SG phase.  
   
   The bound trimers appearing in the SG phase undergo crystallization - phase TC1
   in Fig.~\ref{f.PHD} - for moderate repulsion and extreme mass imbalance. 
   The TC1 phase is analogous to the one discussed in Ref.~\onlinecite{Keilmannetal09}
   (see also Ref.~\onlinecite{SuppMat}).
   This phase is marked by a suppression of  $C_{nn,\bar{b}}$ which attains its absolute
   minimum $C_{nn,\bar{b}} = 1- 2n$ for $j\to 0$, due to the suppression of the weight of 
   configurations with contiguous trimers. The resulting non-monotonic behavior of $C_{nn,\bar{b}}$
   as a function of $j$, shown in Fig.~\ref{f.SG}(b) (see also Ref.~\cite{SuppMat}), provides further evidence that the SG
   phase is a liquid of pre-formed trimers - the enhancement of $C_{nn,\bar{b}}$ in that phase
   is due to trimer binding, while the suppression in the TC1 phase is due to crystal ordering
   of the trimers. 

  Our theoretical results have immediate consequences for current experiments 
  on one-dimensional mixtures of mass-imbalanced cold atoms. Such experiments
  can probe both attractive and repulsive interactions, within the same experimental
  conditions, via the use of Feshbach resonances \cite{Chinetal10}. Our phase
  diagram reveals a fundamental asymmetry between the attractive and the repulsive 
  case for weak and moderate mass imbalance, with the formation of bound pairs on the 
  attractive side and the absence of spin gap on the repulsive one. This asymmetry 
  is very well seen in time of flight experiments probing the momentum distributions
  $n_a(k)$, $n_b(k)$, which are very broad on the attractive side, while they exhibit
  sharp quasi-condensation peaks on the repulsive side -- as shown in Fig.~\ref{f.nk}.
  This asymmetry can be used as strong evidence of pairing on the attractive side.   
  A partial symmetry is recovered only for strong mass imbalance, with the opening
  of a spin gap in the repulsive case, and the occurrence of a crystalline phase for both
  signs of the interaction. This is also well captured by the momentum distributions, 
  showing this time a suppression of the quasi-condensation peaks on the 
  repulsive side due to the appearance of the SG phase and of the crystalline phase.  
  Finally the availability of high-resolution \emph{in-situ} imaging \cite{Bakretal09} 
   sensitive to the spin \cite{Weitenbergetal11} allows to detect the formation of 
   bound trimers and the enhancement of string correlations characterizing
   the SG phase. 
   

We thank P. Azaria, S. Capponi, T. Giamarchi, F. Heidrich-Meisner, and E. Orignac for 
fruitful discussions, F. Ortolani for help with the DMRG code, 
and the PSMN (ENS-Lyon) for generous computer support. 
CDEB is grateful to CNISM Unit of the Physics Department, University of Bologna, where this work was started.

\newpage

\section{Supplementary material}
 
\subsection{Attractive case: PSF-CDW and CDW-crystal transition}

\begin{center}
\begin{figure}[tb]
\includegraphics[width=65mm]{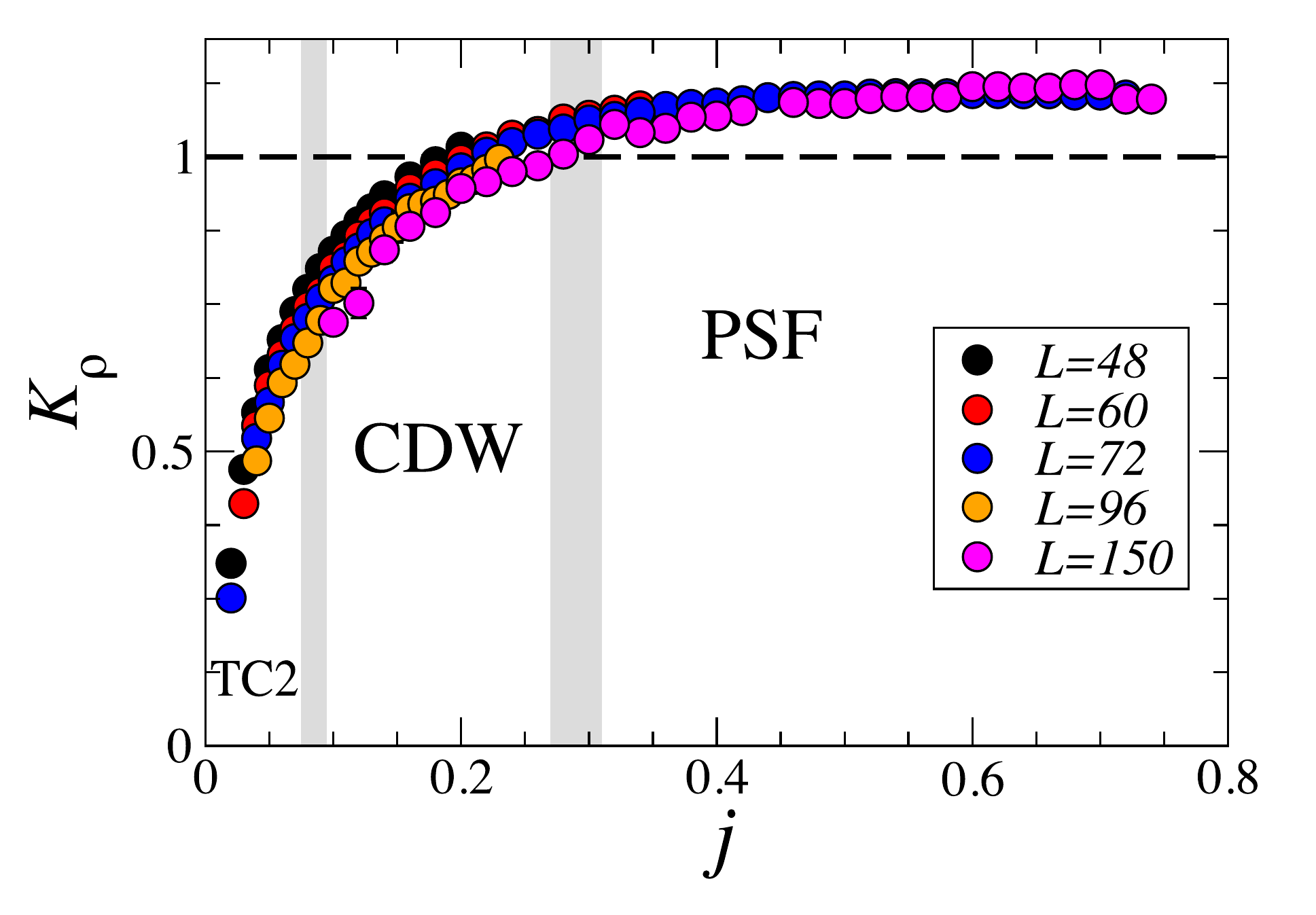}
\caption{Luttinger exponent $K_{\rho}$ as a function of mass imbalance for the 
attractive case $u=-1$ and $n=1/3$. The shaded areas mark the estimated transition
regions between the various phases.}
\label{f.PSF-CDW}
\end{figure}
\end{center}

 In this section we focus on the attractive case $u<0$. The transition from 
 the PSF phase to the CDW phase, obtained for decreasing hopping ratio $j$, 
 is characterized by an inversion of the hierarchy between the pairing correlations, 
 $G_{ab}(r) \sim r^{-1/K_{\rho}}$, and the density-density correlations 
 $C_{\rho}(r) \sim  \cos(2\pi n r) ~r^{-K_{\rho}}$. In the PSF phase $K_{\rho} > 1$, 
 while in the CDW phase $K_{\rho} < 1$. The $K_\rho$ exponent, extracted
 from the low-$q$ behavior of the density structure factor as discussed in the
 text, is shown in Fig.~\ref{f.PSF-CDW} as a function of $j$ for various system sizes. 
 A finite-size extrapolation allows to identify the critical $j$ value at which 
 $K_\rho$ traverses the critical value $K_\rho=1$. 
 The composite $a$-$b$ pairs form therefore a TLL with effective exponent
 $K_{\rm eff} = K_\rho/2$. In the CDW phase, we obtain a TLL with $K_{\rm eff} < 1/2$, which is 
 quite remarkable in a system with on-site interactions only. Indeed in one-component 
TLLs such a situation is realized only in presence of non-local interactions, such as
nearest-neighbor or dipolar ones (R. Citro \emph{et al.}, New J. Phys. 10 , 045011 (2008); 
T. Roscilde and M. Boninsegni, New J. Phys. {\bf 12}, 033032 (2010);  M. Dalmonte \emph{et al.}, 
Phys Rev Lett. {\bf 105}, 140401 (2010)) or when considering highly-excited states such as
a super-Tonks-Girardeau gas (G. E. Astrakharchik \emph{et al.}, 
Phys. Rev. Lett. {\bf 95}, 190407 (2005)).
 In our case, although
 the interactions between the $a$ and $b$ particles are on-site only, the bound $a$-$b$
 pairs have effective longer-range interactions, decaying exponentially with the distance. 

\begin{center}
\begin{figure}[tb]
\includegraphics[width=70mm]{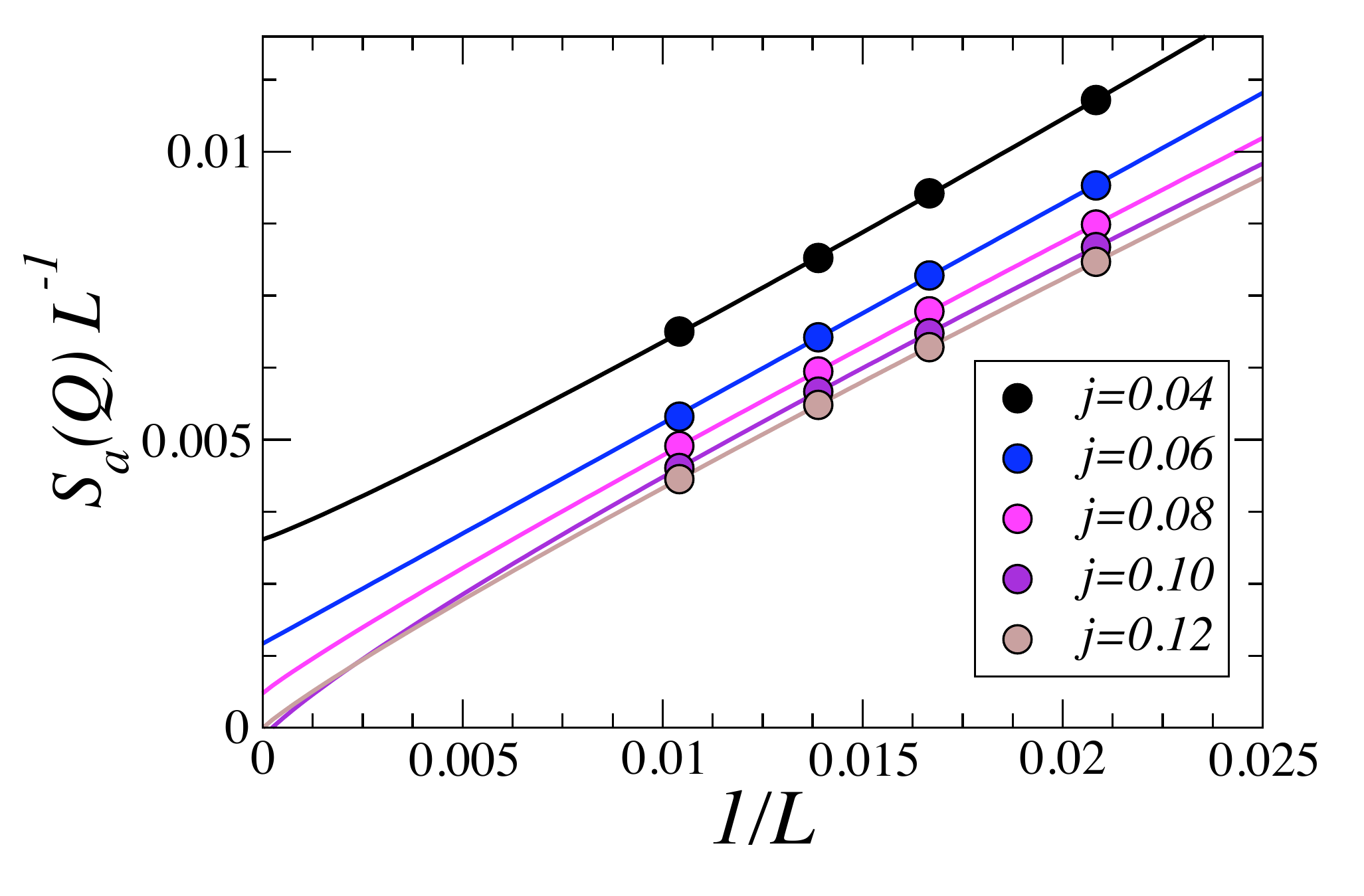}
\includegraphics[width=70mm]{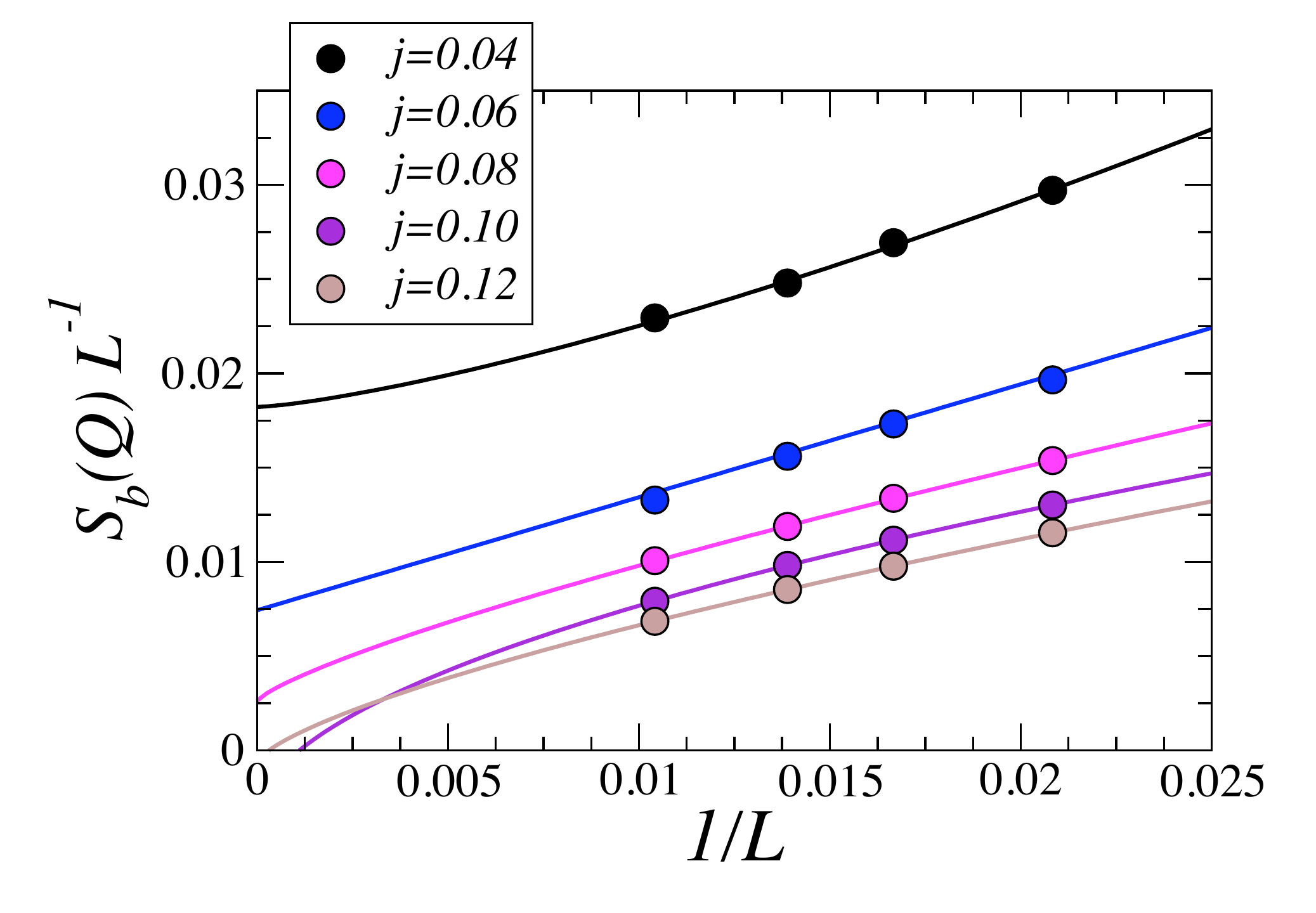}
\caption{Density structure factor for $a$ and $b$ particles at $u=-1$ and various
$j$ values. Solid lines are fits to the form $a_1+a_2/L^{a_3}$.}
\label{f.CDW-TC}
\end{figure}
\end{center}

These effective interactions are responsible for the transition to crystalline order (phase TC2) at 
extreme mass imbalance. This transition occurs at the critical value $K^{(c)}_{\rm eff}=2/p^2$, where
$p = 1/n \in\mathbb{N}$\cite{Giamarchi04}. Since in our case $p=3$, one obtains $K_{\rho}^{(c)}=4/9$ at the transition
line, and  $K_{\rho} = 0$ in the crystalline phase, marking the onset of long-range density-density 
correlations. This implies the appearance of a peak diverging linearly in $L$ at wavevector $Q = 2\pi n$ 
in the structure factor
for the total density $S_{\rho}(q)$, or alternatively in the density structure factor of both
species
\begin{equation}
S_{a(b)}(q) =  \sum_r e^{iqr} \left[ \langle n_{i,a(b)}  n_{i+r,a(b)} \rangle -  \langle n_{i,a(b)} \rangle \langle n_{i+r,a(b)} \rangle \right]~.
\end{equation}
The onset of a linear-in-$L$ divergence of the structure factor peak is shown in Fig.~\ref{f.CDW-TC},
where the peak value divided by $L$ is seen to converge to a finite value in the thermodynamic
limit. This criterion has been used to determine the phase boundary between TC2 and CDW in 
Fig.~\ref{f.PHD}.

\begin{figure}[tb]
\includegraphics[width=65mm]{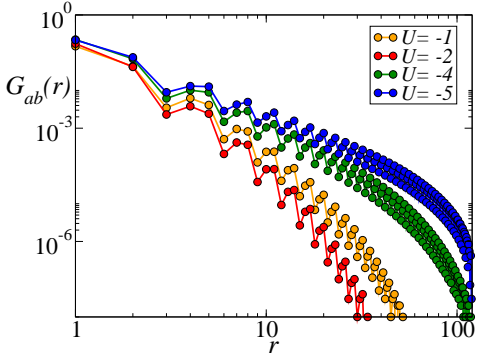}
\caption{Pairing correlation function on a $L=144$ system starting from $L/6$ for $j=0.03$ and 
different interaction strength. }  
 \vspace*{-.4cm}
\label{f.Gab_corr}
\end{figure}

A further clear signature of the TC2 phase is the exponential suppression of the pairing correlator $G_{ab}$,
in sharp contrast with the algebraic decay of both CDW and PSF phases.
We have used DMRG calculations with up to $L=144$ sites for \emph{fermionic} $a$ and $b$ 
species in order to determine
the extent of the TC2 phase (diamonds in Fig.~\ref{f.PHD}). A typical set of data for $j=0.03$ 
is presented in Fig. \ref{f.Gab_corr}: there it is seen that for $u=-1$ and $-2$ the  correlator $G_{ab}$
decays exponentially; interestingly, for \emph{stronger} attraction, \emph{i.e.} for more
tightly bound pairs, the correlation function acquires again an algebraic decay, showing the 
re-entrant nature of the TC2 phase. 
 We observe that, close to the onset of the TC2, the decay of the  $G_{ab}$ correlator
 is indeed consistent with an exponent $K_{\rho}^{(c)}=4/9$.

\subsection{Repulsive case: entanglement entropy analysis}

\begin{figure}[b]
\includegraphics[width=42mm]{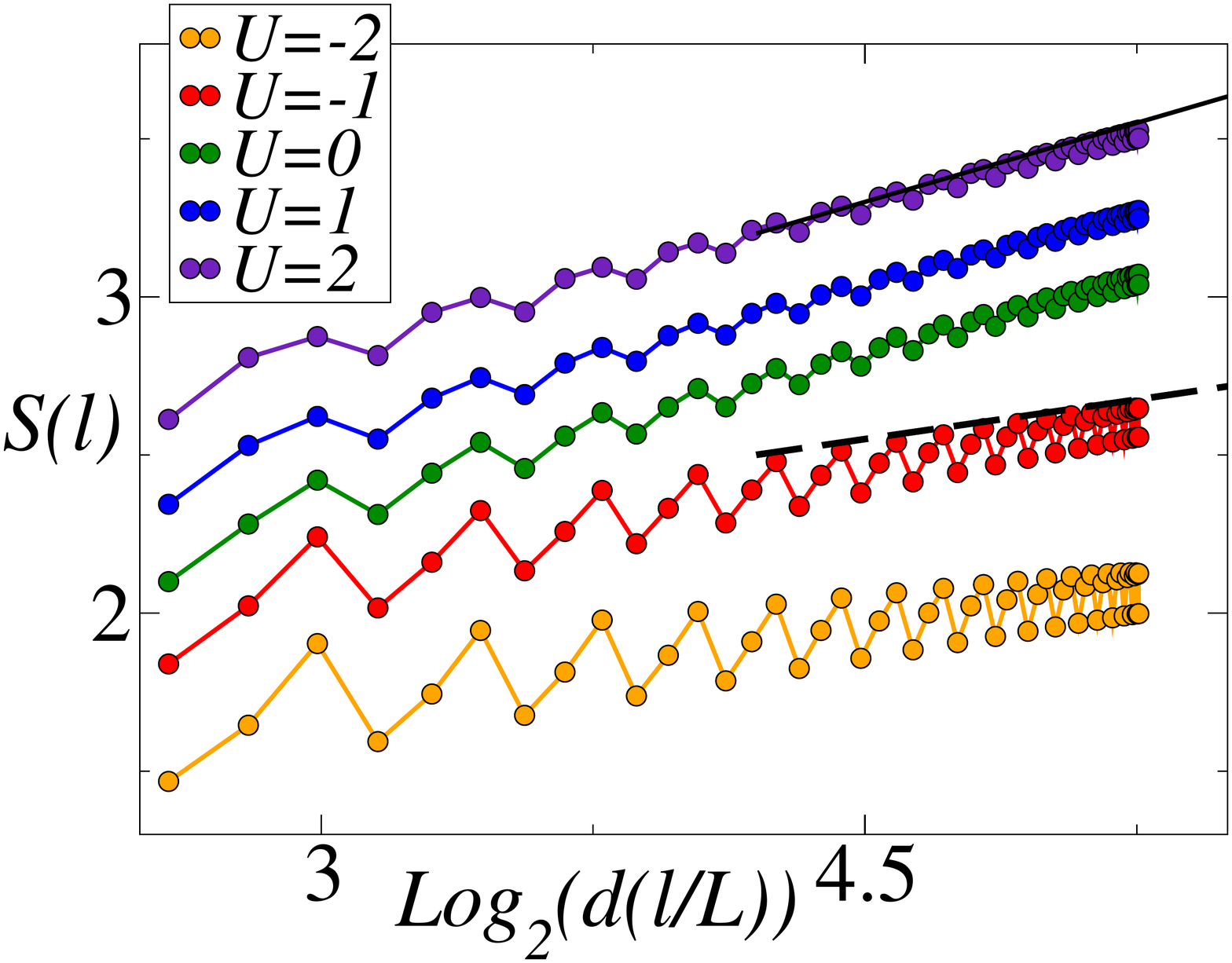}
\includegraphics[width=42mm]{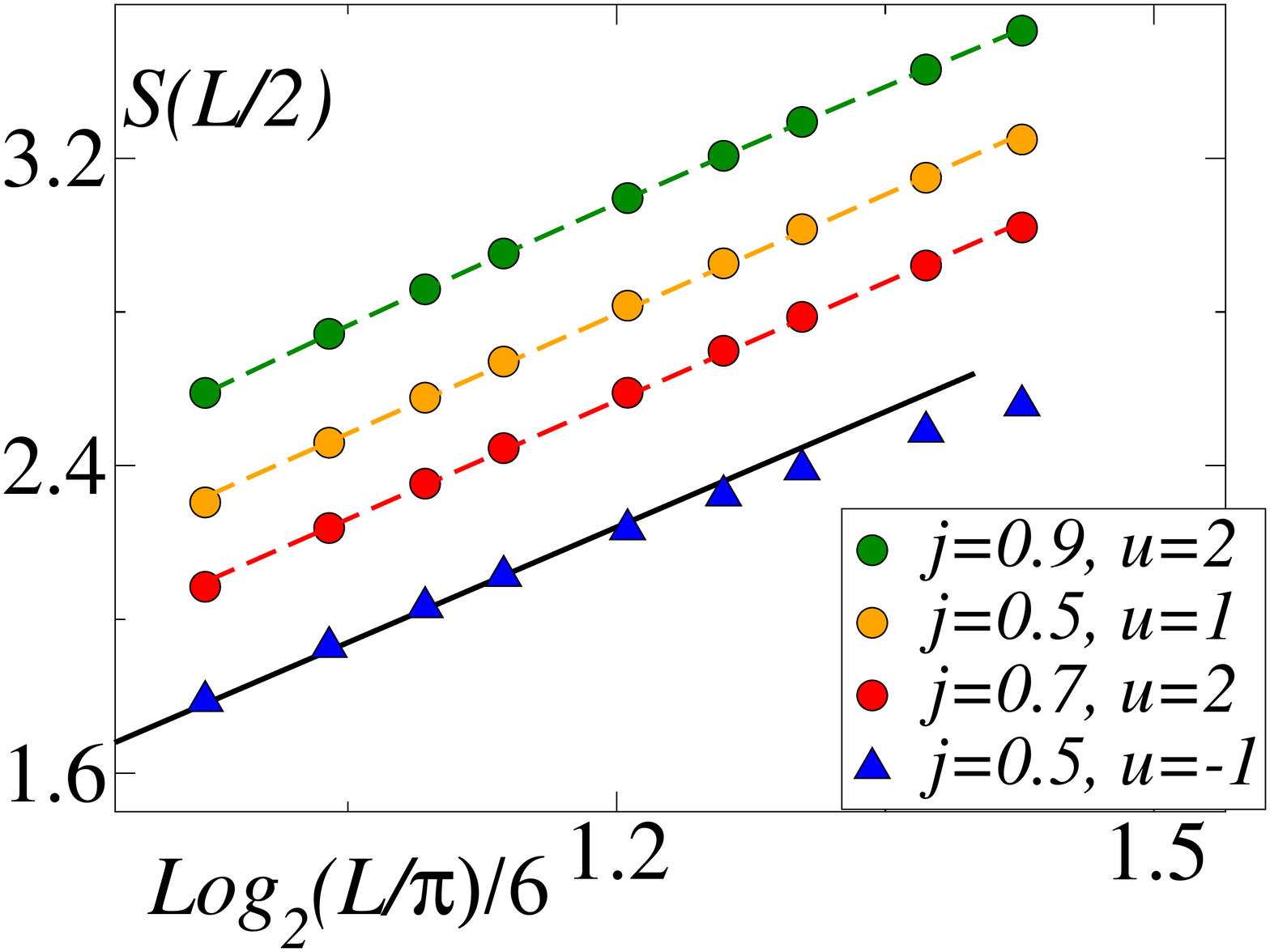}
\caption{Entanglement entropy of a subblock of length $l$. Left panel: EE scaling for $L=120, j=0.5$ and 
different values of $u$; for $u\geq 0$ and $u<0$, results are compatible with $c=2$ (black thick line) and $c=1$ (dashed line) respectively. Right panel: scaling of EE for $l=L/2$ at different system sizes. In the repulsive case, datas are compatible with $c=2$, whereas in the attractive one datas for larger system deviates toward $c=1$ at large $L$; dashed lines are linear fits with $c=2.01, 2.07, 2.05$ from top to bottom, whereas the black, thick line represents $c=2$. Data points for $(j=0.5, u=1)$ and $(j=0.9, u=2)$ have been shifted by $0.2, 0.5$ to improve readability. }  
 \vspace*{-.4cm}
\label{f.EE_sup}
\end{figure}

As mentioned in the main text, the entanglement entropy (EE) allows to extract the central charge of the system\cite{CalabreseC04}. In Fig. \ref{f.EE_sup},
we consider two different approaches to extract $c$. In the left panel, we plot $S(l)$ as a function of the block length $l$ 
at a fixed hopping asymmetry $j=0.5$ and fixed length $L$, for different values of $u$; in the repulsive case, the numerical results are compatible with a completely
gapless system, whereas in the attractive case the emergence of the spin gap reduces the central charge to 1. In the right panel, we
plot the EE of the half chain
\begin{equation}
S(L/2)=\frac{c}{6}\log_2(L/\pi)+ \mathcal{O}(1/L) + {\rm const.}
\end{equation}
which, for $L=6m, m\in\mathbb{N}$, does not exhibit any oscillatory correction. For $u>0$, all datas fit very well with the CFT prediction with $c\simeq 2$, whereas a clear bending at large system sizes indicates that a gap opens in the attractive case.

\subsection{Repulsive case: nearest-neighbor density correlations}
 
\begin{figure}[tb]
\includegraphics[width=60mm]{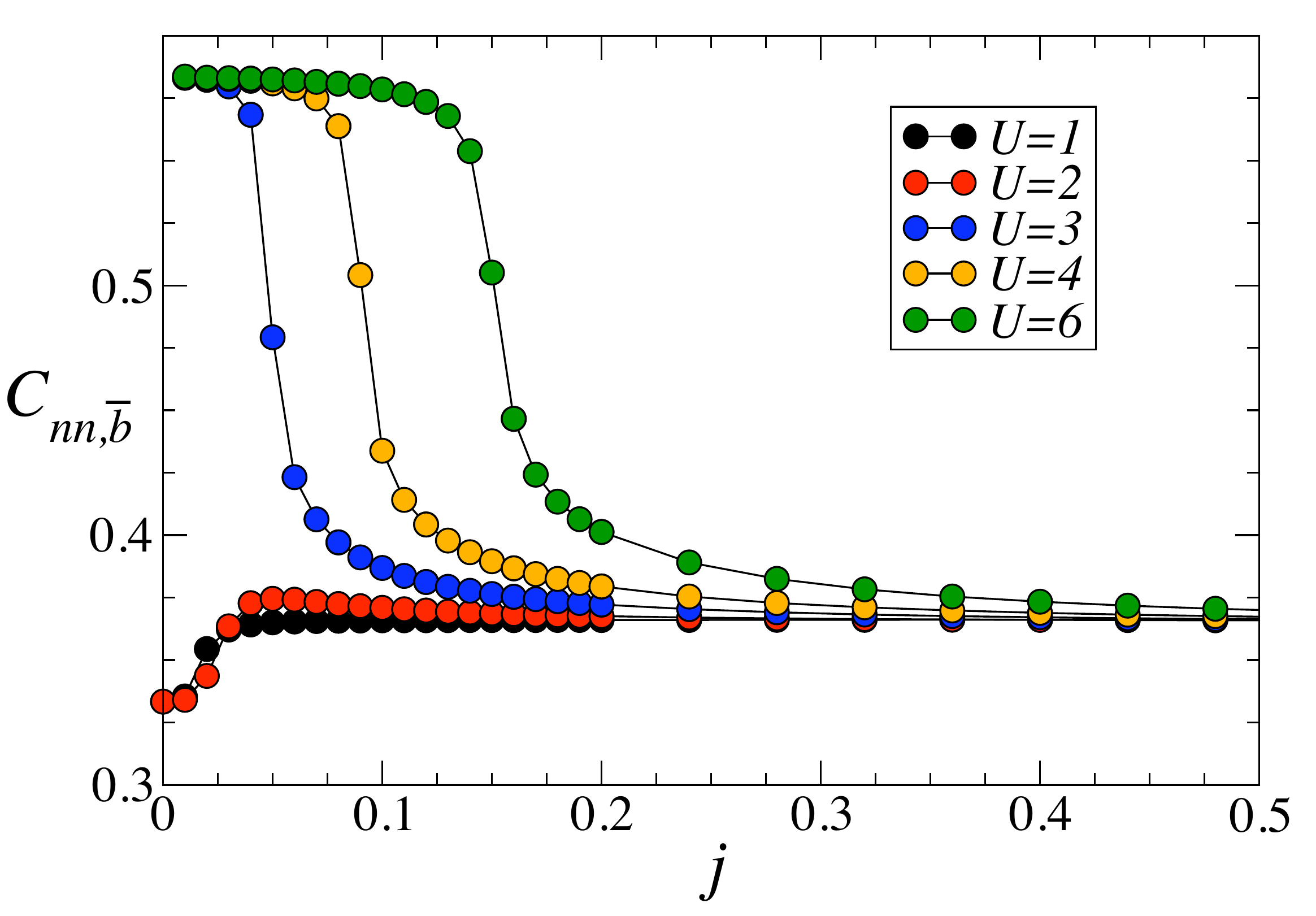}
\caption{Nearest-neighbor correlation function for $b$ holes ($\bar{b}$) for a chain
or $L=12$ sites.}  
\label{f.Cnnbar}
\end{figure}

  Fig.~\ref{f.Cnnbar} shows the nearest-neighbor correlation function 
  for $b$ holes, $C_{nn,\bar{b}}  = \langle (1-n_{i,b})(1-n_{i+1,b})\rangle$
 in the repulsive case. The calculation is performed via exact diagonalization
 on a chain with $L=12$ sites, which allows to access properly the true ground
 state in the limit $j\to 0$, both in the case of crystalline order and of 
 phase separation. The latter limit is hard to reproduce appropriately with 
 QMC or DMRG simulations, due to the abundance of metastable states close
 in energy to the ground state. 
 We see that $C_{nn,\bar{b}}$  is very sensitive to the succession of phases traversed
 by the system. As explained in the text, for small $u$ values $C_{nn,\bar{b}}$ is 
 suppressed as $j\to 0$ due to the formation of the crystal of $a-2\bar{b}$ trimers, 
 implying that $C_{nn,\bar{b}}$ has to attain its minimum value $1 - 2n = 1/3$.
 As $j$ is reduced, two competing effects result in a non-trivial evolution of 
 $C_{nn, \bar{b}}$. The formation of $a-2\bar{b}$ trimers leads to an enhancement
 of the $C_{nn,\bar{b}}$ correlator: the kinetic repulsion between the $\bar{b}$
holes, which would be normally present in a 1D single-component gas with 
hardcore repulsion, is strongly
suppressed due to the decrease of the hopping ratio $j$, and therefore it is
overcome by the binding effect of the light $a$ particles, which gain energy
by delocalizing over two (or more) $\bar{b}$ holes. Yet the trimers that form 
via this mechanism experience an effective mutual repulsion at sufficiently small $u$ values, 
due to the kinetic repulsion between light $a$ particles. This leads to the
suppression of configurations with adjacent trimers; a complete suppression
of such configurations would lead the system to the trimer crystal state, which, 
as mentioned before and in the text, corresponds to an absolute minimum of 
 $C_{nn, \bar{b}}$. Given that trimers are only loosely bound for small $u$, 
 the effect of the repulsion between trimers dominates over the effect of trimer
 formation, so that $C_{nn, \bar{b}}$ is seen to decrease monotonically for 
 decreasing $j$ ($u=1$ data in Fig.~\ref{f.Cnnbar}). On the other hand, for 
 larger $u$, the enhancement of $C_{nn, \bar{b}}$ due to trimer formation
 dominates over the suppression due to trimer repulsion, so that 
 $C_{nn, \bar{b}}$ is seen to increase for decreasing $j$ down to the 
 transition point to the crystalline phase, at which it drops down to the 
 $j\to0$ limit ($u=2$ data in Fig.~\ref{f.Cnnbar}). 
 
 On the other hand, for even larger $u$ the effective attraction between 
  $\bar{b}$ holes mediated by the $a$ particles begins to involve more than
  two holes at once, as the $a$-$b$ interaction overcomes progressively
  the kinetic repulsion between $a$ particles and favors their aggregation over
  extended clusters of holes. This phenomenon culminates into the phase separation
  between hole-rich and hole-free regions\cite{Keilmannetal09, Barbieroetal10}
  upon decreasing $j$ , namely upon suppression of the kinetic repulsion 
  between holes. This succession of phenomena leads to  
  a much stronger enhancement of $C_{nn, \bar{b}}$ as $j$ decreases, with 
  a steep increase when phase separation is achieved. When full phase separation
  is realized in the limit $j\to 0$, 
   $C_{nn, \bar{b}}$ takes the value $[(1-n)L-1]/L$ (namely $7/12 = 0.583333$ 
   for $L=12$, as seen in Fig.~\ref{f.Cnnbar}).

\subsection{Decay of string correlations from bosonization} 

In the following, we will study the dependence on the TLL parameters of the generalized string correlation function:
\begin{equation}\label{corr_string}
\mathcal{O}_{\beta}^z(r)=\left\langle \tilde{S}^z_i ~ \exp\left[i\pi\beta \sum_{i<k<i+r}\tilde{S}^z_k\right]\tilde{S}^z_{i+r}\right\rangle
\end{equation}
where we have defined $\tilde{S}^z_{i}=\frac{n_{a,i}-n_{b,i}}{2}$. After applying standard bosonization identities
to $a, b$ operators by introducing the bosonic $\phi_a,\phi_b$ fields\cite{Giamarchi04}, one can define {\it effective} spin and charge 
fields as $\phi_{\sigma,\rho}=(\phi_a\mp\phi_b)/\sqrt{2}$ such that:
\begin{equation}\label{sz}
\tilde{S}^z_r=\frac{\partial_r\phi_{\sigma}(r)}{\sqrt{\pi}}+\gamma V_{\rho}^{\pm\sqrt{4\pi}}(r)V_{\sigma}^{\pm\sqrt{4\pi}}(r)+...
\end{equation}
where $V_{\sigma,\rho}^{\pm\alpha}(r)=e^{\pm i\alpha \phi_{\sigma,\rho}(r)}$ are vertex operators related to the charge and spin fields and $\gamma$ is a constant; additional contributions with higher scaling dimension play no significant role in the remaining, and can thus be neglected. The non-local contribution in  Eq. \ref{corr_string} can be reduced to: 
\begin{eqnarray}
 \exp[i\pi\beta \sum_{i<k<i+r}\tilde{S}^z_k] &=& \exp[i\pi\beta \int^{i+r}_i dy \frac{\partial_y\phi_{\sigma}(y)}{\sqrt{\pi}}]=\nonumber\\
 &=& \exp[i\sqrt{\pi}\beta [\phi_{\sigma}(i+r)-\phi_{\sigma}(i)]]\nonumber.
\end{eqnarray}
In addition to a pure spin contribution which stems from the first term in Eq.~\eqref{sz}, and which decays faster than $r^{-2}$, 
the string correlator contains a more complicated, second contribution, stemming from vertex operators, and which we will indicate as
 $\mathcal{O}_{\beta}^{z,2}$. Assuming spin-charge separation, $\mathcal{O}_{\beta}^{z,2}$ can be written as:
\begin{eqnarray}\label{stringz}
\mathcal{O}_{\beta}^{z,2}(r)&\propto& \langle V_{\rho}^{+\sqrt{4\pi}}(i)V_{\rho}^{-\sqrt{4\pi}}(i+r)\rangle \times\nonumber\\
&\times& \langle V_{\sigma}^{\pm\sqrt{4\pi}}(i)V_{\sigma}^{-\beta\sqrt{\pi}}(i)V_{\sigma}^{\beta\sqrt{\pi}}(i+r)V_{\sigma}^{\mp\sqrt{4\pi}}(i+r) \nonumber\\
&=& \mathcal{M}_{\sigma}^{(\beta)}(r)* (1/r^{K_{\rho}}).\nonumber
\end{eqnarray}
whose asymptotic decay is affected by both charge and spin contributions: the charge part has a fixed exponent, whereas the spin one $\mathcal{M}_{\sigma}^{(\beta)}(r)$ has a $\beta$-dependent form; in particular, for $\beta=2$ correlation considered in this work, the slowest decaying part of $\mathcal{M}_{\sigma}^{(\beta)}(r)$ is a constant, so that the dominant term in the string correlation functions decays as:
\begin{equation}
\mathcal{O}_2^z(r)\propto 1/r^{K_{\rho}}
\end{equation}
It is worth noticing that, from Eq.~\eqref{stringz}, one can easily recover standard spin correlations for $\beta=0$ decaying as $1/r^{K_{\rho}+K_{\sigma}}$.
 
\begin{figure}[tb]
\includegraphics[width=60mm]{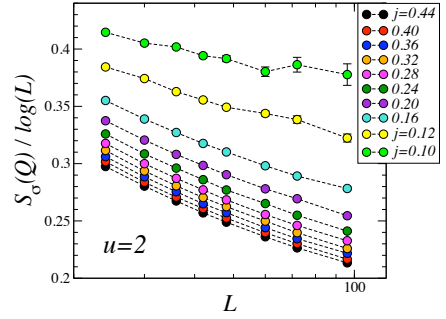}
\caption{Scaling of the peak in the spin structure factor for  the repulsive case, $u=2$.}  
\label{f.SQ}
\end{figure}

The fact that the decay of string correlations is uniquely governed by the $K_{\rho}$
exponent is indeed remarkable; as discussed in the main text, this leads
to a string structure factor diverging with system size as 
$O_{z,int} \sim L^{1-K_{\rho}}$ (and $K_{\rho} < 1$ for all $j$ values). 
This result has to be compared with the scaling of the peaks of the 
charge and spin structure factor, namely for $S_{\rho}(Q)$ and $S_{\sigma}(Q)$ 
with $Q = 2\pi n$. These peaks come from the dominant oscillating term 
(going like $\cos(Qr)$) of  the $C_{\rho}(r)$ and $C_{\sigma}(r)$ correlation functions, whose
amplitude decays like $r^{-K_{\rho}-K_{\sigma}}$. This means that in general
the $Q$-peak height scales with system size as $L^{1-K_{\rho} - K_{\sigma}}$. 
In the SG phase the $K_\sigma$ exponent vanishes for $L\to\infty$, 
so that, strictly speaking, the true asymptotic scaling of  $S_{\rho}(Q)$ and $S_{\sigma}(Q)$ is 
indeed the same as that of $O_{z,int}$. But \emph{in practice} the scaling of 
$K_{\sigma}(L)$ to zero is extremely slow -- typically logarithmic, 
a fact which generates a huge finite-size correction to the effective scaling of 
$S_{\rho}(Q)$ and $S_{\sigma}(Q)$. In our simulations we observe that 
these quantities grow more slowly than $\log(L)$ over the entire SG phase for all the 
system sizes we considered (up to $L=96$ -- see Fig.~\ref{f.SQ}). 
To see them diverge at all with system size one might need exceedingly big system sizes, both for 
simulations and for experiments. Therefore string correlations stand as a 
unique tool to characterize the correlations in the SG phase. 

\begin{center}
\begin{figure}[tb]
\includegraphics[width=70mm]{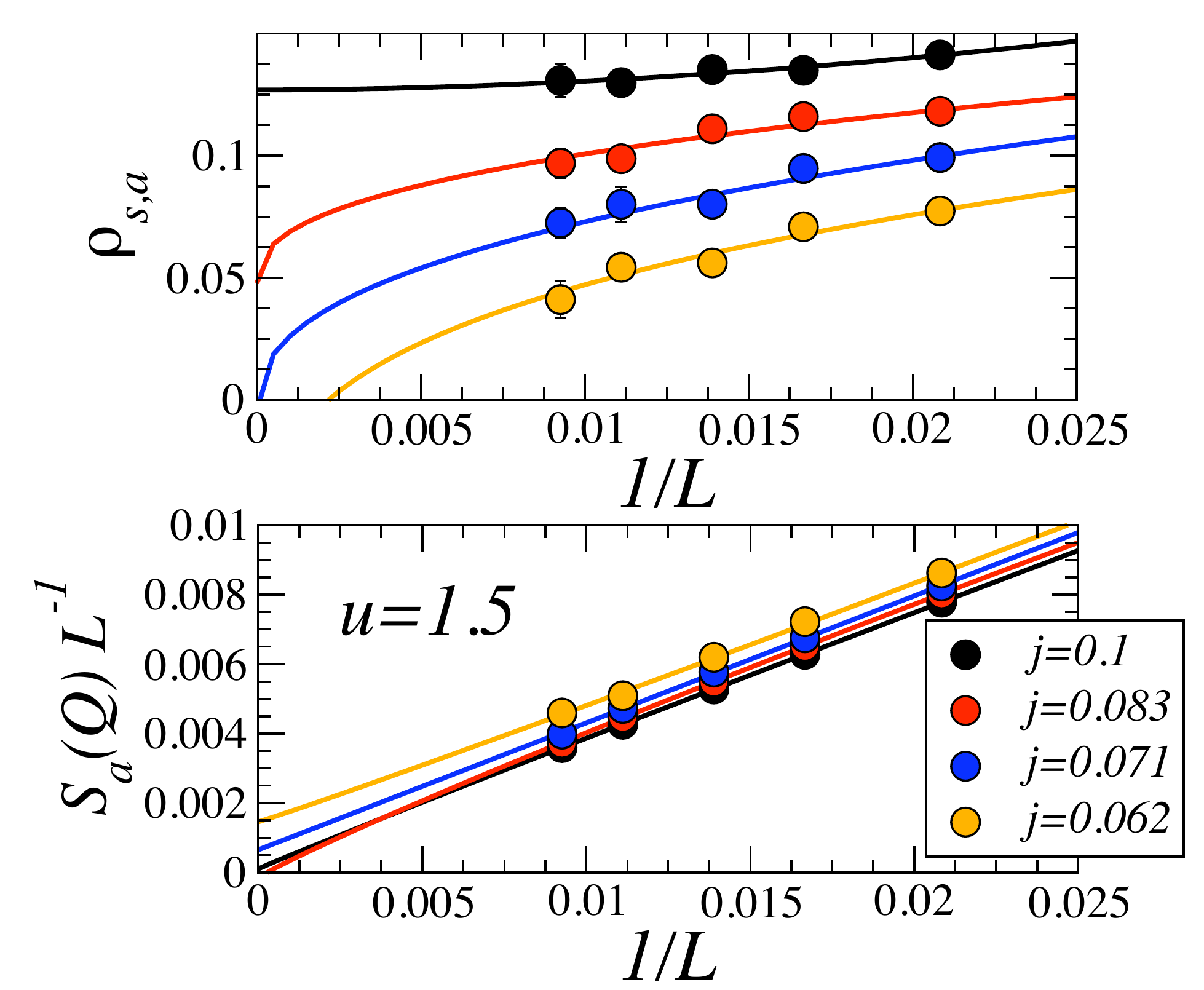}
\caption{Upper panel: scaling of the superfluid density of 
the $a$ particles for $u =1.5$. Lower panel: scaling of the  
peak in the density structure factor for the $a$ particles. All solid lines are fits to the form $a_1+a_2/L^{a_3}$.}
\label{f.SG-TC1}
\end{figure}
\end{center}

\subsection{Repulsive case: onset of the crystalline phase}

 In this section we provide further details of the estimate of the transition line between the
 SG and the TC1 phase in the repulsive case. The transition is obtained by a joint finite-size
 scaling analysis. We focus on the scaling of the peak in the structure 
 factors $S_{a(b)}(q)$ appearing at $Q=2\pi n$, and diverging linearly with system size 
 when entering the TC1 phase; and we consider as well the scaling of the superfluid
 density for both species, $\rho_{s,a(b)}$, which scales to zero in the thermodynamic
 limit when in the TC1 phase. Fig.~\ref{f.SG-TC1} shows that the scaling of 
both quantities give a consistent estimate of the critical $j_c$ value for the onset of 
crystalline order. 

The transition to the TC1 phase was already estimated by one of the authors in 
Ref.~\cite{Keilmannetal09} for the same density. The estimate of the transition
line we present here is corrected with respect to that given in Ref.~\cite{Keilmannetal09}.
Indeed in Ref.~\cite{Keilmannetal09} the estimate of $j_c$ was biased by the
choice of a scaling ansatz on the superfluid stiffness which, albeit satisfactorily
verified within the statistical uncertainties, turns out to not be fully 
justified. The current estimate is not based on any scaling ansatz, and it delivers 
a transition line which is shifted to larger $j_c$ values (typically by about 10-20 \%)
 with respect to the estimates in Ref.~\cite{Keilmannetal09}.
 
Moreover Ref.~\cite{Keilmannetal09} identified a narrow region of parameters close to 
the transition line as a \emph{super-Tonks} regime, characterized by a sub-linearly
diverging peak in the density structure factor of both species. The subtle effect of the opening 
of a spin gap before entering the crystalline phase 
was not observed in Ref.~\cite{Keilmannetal09}, and this led to the erroneous claim of algebraic 
off-diagonal correlations for both species in the super-Tonks regime.  
Our current results point to the fact that the above regime is rather characterized by 
superfluidity of both atomic species (which persists up to the transition, as shown in 
Fig.~\ref{f.SG-TC1}) but with exponentially decaying off-diagonal correlations, due to the
presence of a spin gap.

\end{document}